\documentclass[twoside,9pt]{article}

\usepackage{arxiv}
\usepackage[utf8]{inputenc} 
\usepackage[T1]{fontenc}    
\usepackage{hyperref}       
\usepackage{url}            
\usepackage{booktabs}       
\usepackage{amsfonts}       
\usepackage{nicefrac}       
\usepackage{microtype}      
\usepackage{lipsum}

\usepackage{extsizes}
\usepackage[super,sort&compress,comma]{natbib} 
\usepackage[version=3]{mhchem}
\usepackage{mathptmx}
\usepackage{sectsty}
\usepackage{graphicx} 
\usepackage{lastpage}

\usepackage{booktabs}
\usepackage[normalem]{ulem}
\useunder{\uline}{\ul}{}

\usepackage{threeparttable}
\usepackage[version=3]{mhchem} 
\usepackage{amsmath}
\usepackage{siunitx}

\makeatletter
    \providecommand\barcirc{\mathpalette\@barred\circ}
    \def\@barred#1#2{\ooalign{\hfil$#1-$\hfil\cr\hfil$#1#2$\hfil\cr}}
    \newcommand\stst{^{\protect\barcirc}}
    \makeatother

\usepackage{fancyhdr}
\usepackage{fnpos}
\usepackage[english]{babel}
\addto{\captionsenglish}{%
  
}
\usepackage{array}
\usepackage{droidsans}
\usepackage{charter}
\usepackage[T1]{fontenc}
\usepackage[usenames,dvipsnames]{xcolor}
\usepackage{setspace}
\usepackage{hyperref}
\newcommand\bfootnote[1]{%
  \begingroup
  \renewcommand\thefootnote{}\footnote{#1}%
  \addtocounter{footnote}{-1}%
  \endgroup
}

\title{Nano-Pump based on Exothermic Surface Reactions}

\author{
  Shaltiel Eloul\\Department of Chemistry\\
  University of Cambridge, UK.\\
  se399@cam.ac.uk
  \And
  Daan Frenkel\\Department of Chemistry\\ 
  University of Cambridge, UK.\\
  df246@cam.ac.uk
}

\begin{document}
\twocolumn[
  \begin{@twocolumnfalse}

\maketitle

\begin{abstract}
We present simulations indicating that it should be possible to construct a switchable nano-scale fluid pump, driven by exothermic surface reactions. Such a pump could, for instance, be controlled electro-chemically. In our simulations we explore a simple illustration of such a pump. We argue that the simplicity of the pump design could make it attractive for micro/nano-fluidics applications.
\end{abstract}

  \end{@twocolumnfalse}
  ]
\section{Introduction}
Inducing flow in a nano-channel requires a mechanism which can efficiently convert energy input into a sustained fluid flow.
As pressure-driven flow becomes less efficient in sub-micron channels, there is much interest in pumps that use electro~\cite{Takamura,Santiago}, diffusio ~\cite{Ajdari_osmosis, DifMichelin, LeeExpOsmosis} or thermo ~\cite{thermoRipoll,LiuThermo} osmosis transport mechanisms. 
Electro-chemical micro-pumps, which generate convective flow~\cite{KlineECPump} are also being explored, but require relatively large channels (order of $10-100\mu m$) that can sustain fully developed convective flows.  

\begin{figure}[!ht]
\centering
\includegraphics[width=0.9\columnwidth]{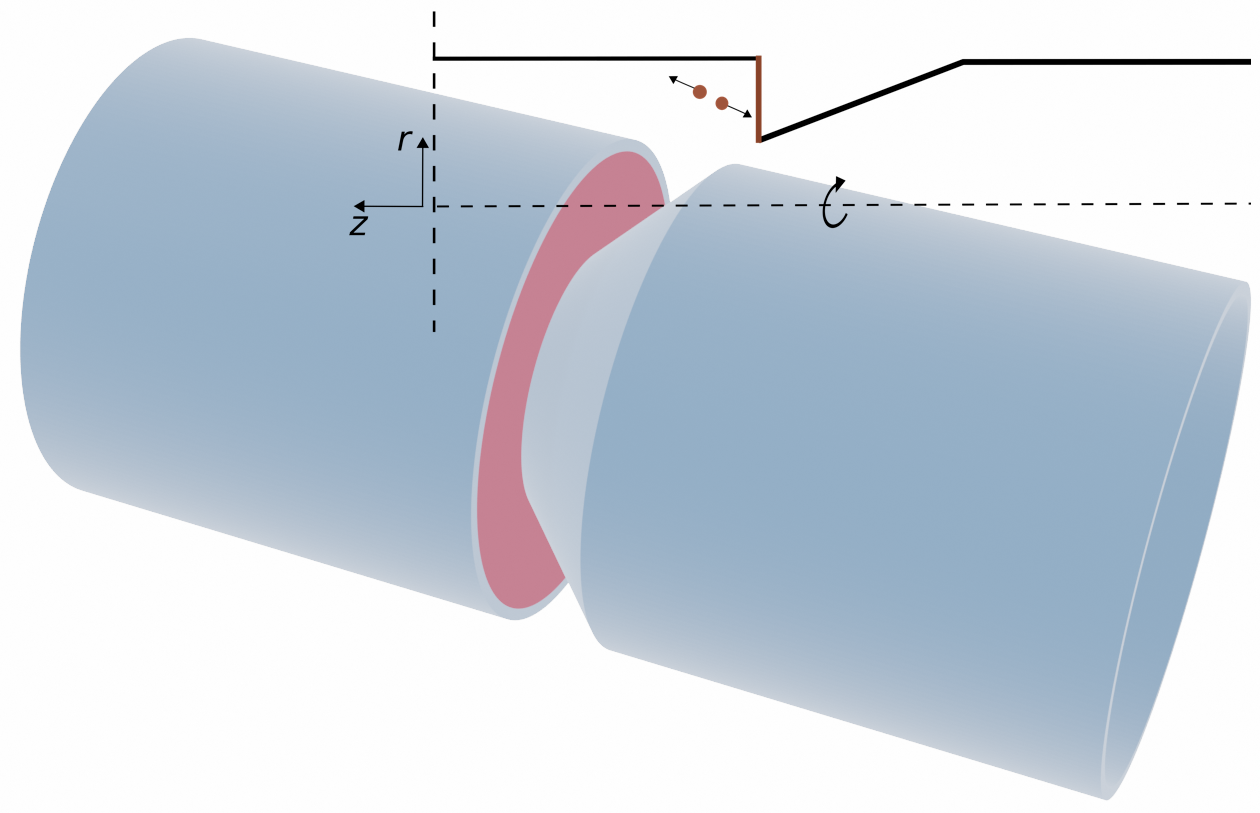}
\caption{Sketch of the nano-pump described  in the text. The active element of the pump is a reactive (e.g. electrode) surface perpendicular to the tube axis.  
The insert (top right) in the figure shows a cut through the channel, parallel to the $z$-direction. 
The drawing shows a pair of fluid particles that have picked up equal and opposite momenta due to the exothermic surface reaction. Part of the momentum is absorbed by the reactive surface, the rest drives the fluid flow.}
\label{geometry}
\end{figure}

Here, we propose that robust fluidic pumping, can be based on locally switchable electrochemical or catalytic reactions that release energy on a reactive surface.   
Our approach is based on our recent observation that exothermic energy released on a catalytic platinum surface can propel micro-sized Janus particles at a speed of a few microns per second~\cite{Rocket}. 
If, instead of considering an uncontrolled chemical reaction on a Janus particle, we consider a reaction on suitably oriented reactive surface (e.g. on fixed electrodes), then the surface reaction could  generate flow by exploiting the large, directional momentum release into the fluid at the electrode surface (see Fig.~\ref{geometry}). 
In the case of electrochemically driven flow, the  exothermic energy release is driven by applying over-potential that converts an electro-active redox species in solution.  
A representative geometry of such a reaction-driven pump is shown in Fig.~\ref{geometry}. In this pump geometry, we assume a cylindrical channel with an asymmetric conical constriction: the active surface is chosen perpendicular to the tube axis, while the conical surface is inert.  

We use a general particle-based model (dissipative particle dynamics, DPD ~\cite{hoogerbrugge1992dpd,espanol1995}) to describe the effect of a local kinetic energy release on the microscopic flow properties of a fluid.   
DPD provides a cheap, yet realistic description of micro-scale hydrodynamic flow in compressible fluids~\cite{groot1997dissipative}. 
In principle, the surface reaction might also drive thermo-osmotic and diffusio-osmotic flows (provided the surfaces are suitably functionalized). 

Here, we use a DPD model, as it allows us to focus on the effect of direct momentum transfer: in this model we can switch off thermal gradients to neglect thermo-osmosis, and by using a one-component fluid, we eliminate possible diffusio-osmosis. It is noted that interaction wall-fluid has been chosen to be repulsive such that there is negligible excess enthalpy near the surface and thermo-osmosis is expected to be negligible in the DPD model even with no thermostat.  

In DPD, the interaction between fluid particles is described as a sum of pair forces: 
\begin{eqnarray}
\bf{f_i} =\sum_{j\neq i}{(F_{ij}^C + F_{ij}^{D} + F_{ij}^R)}\;,
\label{eqdpd}
\end{eqnarray}
where $\bf{F^{C,D,R}}$ denote respectively the conservative, dissipative and random components of the force. 
The pair interaction between fluid particles is given by a quadratic, repulsive potential, resulting in a pair force:
\begin{eqnarray}
{\bf{F_{ij}^C}} =\alpha (1- \frac{r_{ij}}{r_c})\bf{\hat {r}_{ij}}
\end{eqnarray}
where $r_{ij}$ denotes the distance $\bf{|r_i-r_j|}$ and ${\bf \hat {r}_{ij}}={{\bf (r_i-r_j)}}/r_{ij}$ is the unit vector in the direction of $\bf{|r_i-r_j|}$. The constant $\alpha$ sets the strength of the repulsion: it is chosen to mimic the compressibility of water~\cite{groot1997dissipative}. 
The dissipative and random forces are related via the fluctuation-dissipation relation~\cite{espanol1995}:
\begin{eqnarray}
{\bf {F_{ij}^{D}}}= -\gamma \omega (r_{ij}) (\bf {v_{ij}\cdot \hat{r}_{ij})\hat{r}_{ij}}, \nonumber\\
{\bf F_{ij}^{R}}= \sqrt{2\gamma kT \omega (r_{ij})}\frac{dW_{ij}}{dt}\bf{\hat{r}_{ij}},
\label{eq:one}
\end{eqnarray}
where $\bf{v_{ij}=\left(v_i - v_j\right)}$ is the relative velocity, $\gamma$ is the friction coefficient controlling energy dissipation into the fluid, and $W_{ij}$ is a Wiener process: $\int_0^{\Delta t} dW_{ij}=\sqrt{\Delta t}\zeta_{ij}$, where $\zeta_{ij}$ is a standard Gaussian random number. 
The weight function $\omega(r)$ is assumed to be of the form 
$ \omega(r_{ij}) = \left(1-\frac{r_{ij}}{r_c}\right)^2$.

As we chose our DPD model parameters to correspond to water~\cite{groot1997dissipative} (see also Table \ref{tab:table1}), our results should be indicative of pumping rates in aqueous medium, once we correct for the difference between the viscosity of real water and that of the DPD model~\cite{groot1997dissipative}$^{,\dag}$. \bfootnote{$^\dag$ See also the provided supporting information.}
In what follows, we use the cut-off distance $r_c$ as our unit of length: for water,  $r_c \sim \SI{6.46}{\angstrom}$. 
The DPD density is chosen to be $\rho  r_c^3=3.0$. 
To reproduce the compressibility water at room temperature, we choose $\alpha=25$ and $\gamma=4.5$ as proposed by Groot and Warren~\cite{groot1997dissipative}.  
We choose the thermal energy $k_BT$ as our unit of energy, and $m$, the dimensionless mass of a fluid particle is our mass unit. 
As a consequence, our unit of time is $r_c\sqrt{mk_BT}$. 
The equations of motion were integrated using a modified velocity-Verlet algorithm~\cite{groot1997dissipative}  with a time step of $10^{-3}$, corresponding to $3 ps$ in SI units, allowing us to simulate local transient momentum dissipation up to hydrodynamic timescales.
 
\begin{table}
\caption{\label{tab:table1} Simulation parameters of DPD fluid. }
\begin{tabular}{lcr}
Parameter &DPD values&Phys. units\\
\hline
Mass ($3H_2O$) & $m$=1 & $9\cdot 10^{-26}kg$\\
$r$ & $r_c=1$ & $6.46\AA$\\
$\epsilon$ & 1 & $kT$\\
$v$ & $\sqrt{\epsilon/m}$ & $214.2m/s$\\
$t$ & $v/r$ & $3.0ps$\\
$\rho$ & 3.0 &\\
DPD: $\alpha,\gamma$ & 25.0, 4.5 & \\
Q  & $N_f/t$ & $3H_2O/ps$\\

\end{tabular}
\end{table}

The inner surface of the tube geometry is modeled with `frozen' particles and we consider here two different modifications to the surface$^\dag$. 
The first modification allows us to change the hydrodynamic boundary conditions on the tube surface from slip to non-slip.  
In the slip case, the surface is formed using smooth high density of frozen particles to suppress hydrodynamic drag. 
In the non-slip case, we construct the boundary to be a non-smooth surface by corrugating the tube surface sufficiently to measure zero slip velocity of the fluid adjacent to the surface. 
The second modification is that we allow collisions of fluid with the pump surface to be either elastic or inelastic. 
When using only repulsive force (but no friction) in fluid-surface interaction, collisions are elastic and the wall does not act as an energy sink. 
Alternatively, we can use dissipative forces between the fluid and wall particles to define an energy absorbing (or, more precisely, thermalizing) boundary. The energy-absorbing surface is not an ideal inelastic wall,  and some slip can still happen in smooth surfaces. However, high energy from reactions are sufficiently suppressed.

We assume that, during a surface reaction, a significant fraction of the reaction enthalpy is taken up by the fluid. 
For example: the catalytic reaction \ce{2H_2O_2 -> O_2 + 2H_2O} on a \ce{Pt} surface has a standard enthalpy $\Delta H\stst = \SI{1.017}{\electronvolt}$. 
In the case of reversible electrochemical reactions, the heat can be associated with the electrode Peltier heat ($\Pi$)~\cite{BOUDEVILLE}. 
$\Pi$ is related to the apparent enthalpy of change in the reaction: $\Pi-W_e=\Delta H^\phi$.
For example, for the redox couple $\mathrm{[Fe(CN)_6]^{-3/-4}}$, $\Pi \sim \SI{0.5 }{\electronvolt}$ and $\Delta H^\phi \sim 1-\SI{1.5}{\electronvolt}$ ~\cite{zhengeph}. 
Therefore, in our model calculations, we consider energy releases in the range of 0-\SI{2}{\electronvolt}.

The energy $E$ released by a surface reaction into the fluid, is assumed to increase the kinetic energy of relative motion of a pair of neighboring DPD particles close to the catalytic surface. 

As we only change the relative motion of two DPD particles, the total momentum in the fluid is conserved and all species are unchanged. 
As a result of a reaction, the kinetic energy of the DPD particles involved is changed:
\begin{equation}\label{eq:quadr}
{
\Delta {\bf v_a}^2+\Delta {\bf v_a} \cdot ({\bf v_{a}}-{\bf v_{b}}) =\frac{E}{m_f}.
}
\end{equation}
where $\bf{v_a}$ and $\bf{v_b}$ are the particle velocities before the reaction, while $\Delta \bf{v}_a$ and $\Delta \bf{v}_b$ are the velocity changes due to energy injection. 
Momentum conservation implies $\Delta {\bf v_a} = - \Delta {\bf v_b}$. 
The direction of $\Delta {\bf v_a}$ is chosen from a uniform distribution, and its magnitude is calculated from Eqn~ \ref{eq:quadr}.

\section{Results and discussion}
\begin{figure}[h!]
\centering
\includegraphics[width=1.0\columnwidth]{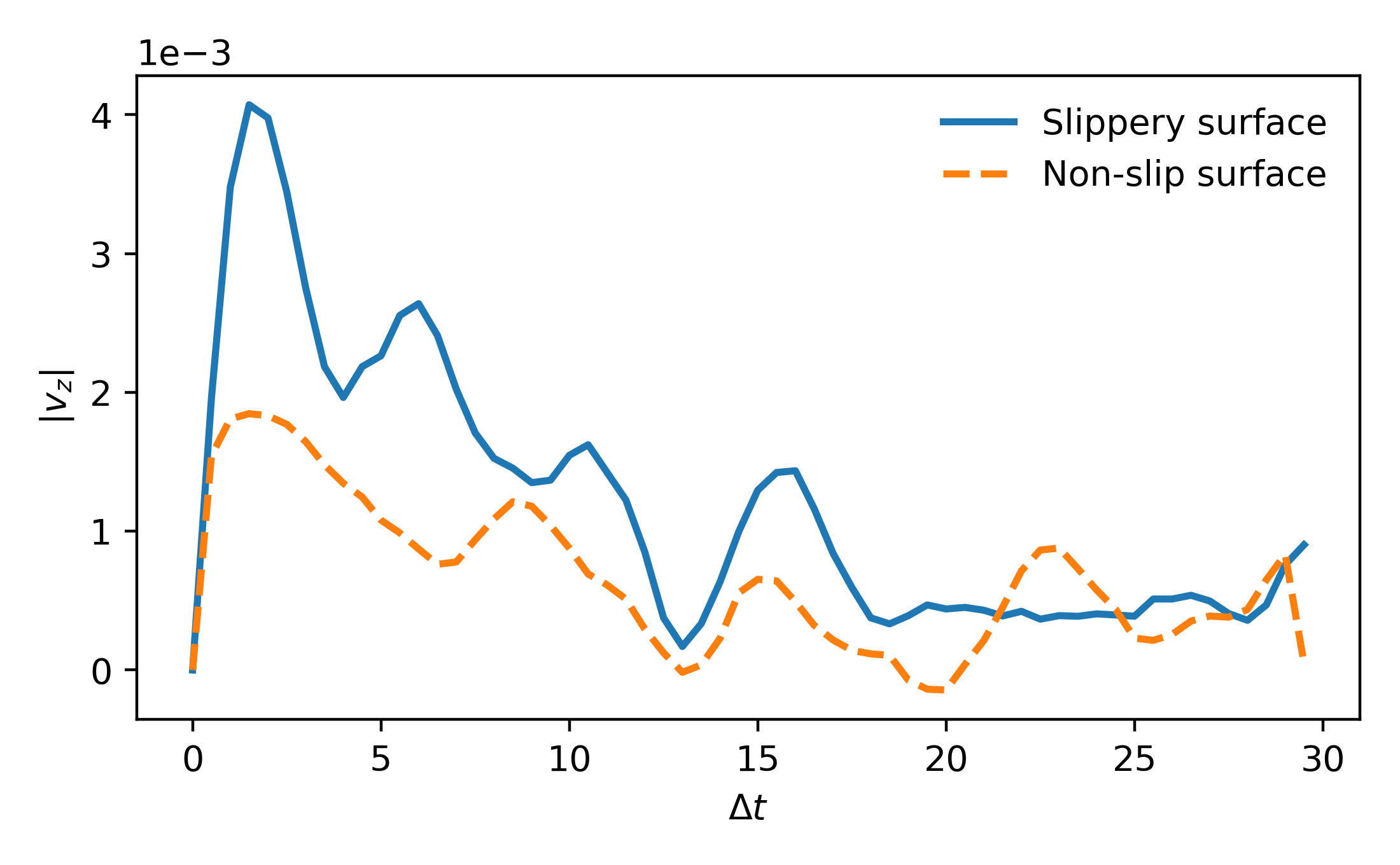}
\caption{$z$-direction velocity transient, obtained by averaging over {\bf 10$^4$} single reactive events, in a pump with radius of 4$r_c$ and a reaction energy of $\SI{1}{\electronvolt}$. 
The figure shows the results for two limiting cases: slip and non-slip boundary conditions on the wall of the channel. In the non-slip case, we also assume inelastic fluid-wall collisions.}
\label{reaction_profile}
\end{figure}

The average time-dependence of the fluid velocity profile from a $\SI{1}{\electronvolt}$ single reaction in a cylinder with radius of 4$r_c$ is plotted in Fig. \ref{reaction_profile}. 
The profile is computed by averaging independent DPD configurations. 
To improve the statistics of the analysis of the surface effect on the momentum decay,  the fluid thermostat was switched off in this specific simulation. 

We show the results of two limiting cases. One case (drawn curve) corresponds to slip and elastic boundary conditions.
The other case corresponds to non-slip boundary conditions, and inelastic fluid-wall collisions  (dashed curve). 
Interestingly, there is an appreciable flow response, even for the case of inelastic, non-slip boundaries.

It is likely that experiments will be closer to the non-slip/inelastic case, although increasing the hydrophobicity of the walls of the channels may change this picture somewhat \cite{Huang}.
 
 The reaction also creates a damped sound wave in the tube (\href{https://www.dropbox.com/s/ak9xv37i4mxrdcg/WEO_anime_momentum.mov?dl=0}{see supporting video}), which results in temporal oscillations of the fluid momentum, but this oscillatory flow field appears not to contribute to the net fluid flow. 
 The fluid momentum imparted by a single reaction decays within 20-30 time units for the non-slip boundary, but for the slip boundary the initial momentum has not yet fully decayed on this time scale. 
 Converting to SI units, transient fluid velocity in the channel peaks at  $\sim 1m/s$ and decays within $100ps$.
Hence, a single reaction event leads to a DPD fluid volume displacement of order $1 \AA$. By scaling the viscosity to water$^\dag$, we expect an order of magnitude smaller displacement from a single event, but multiple reaction events can generate significant pumping rates. 

\begin{figure}[h!]
  \begin{tabular}{@{}c@{}}
     \includegraphics[scale=0.6]{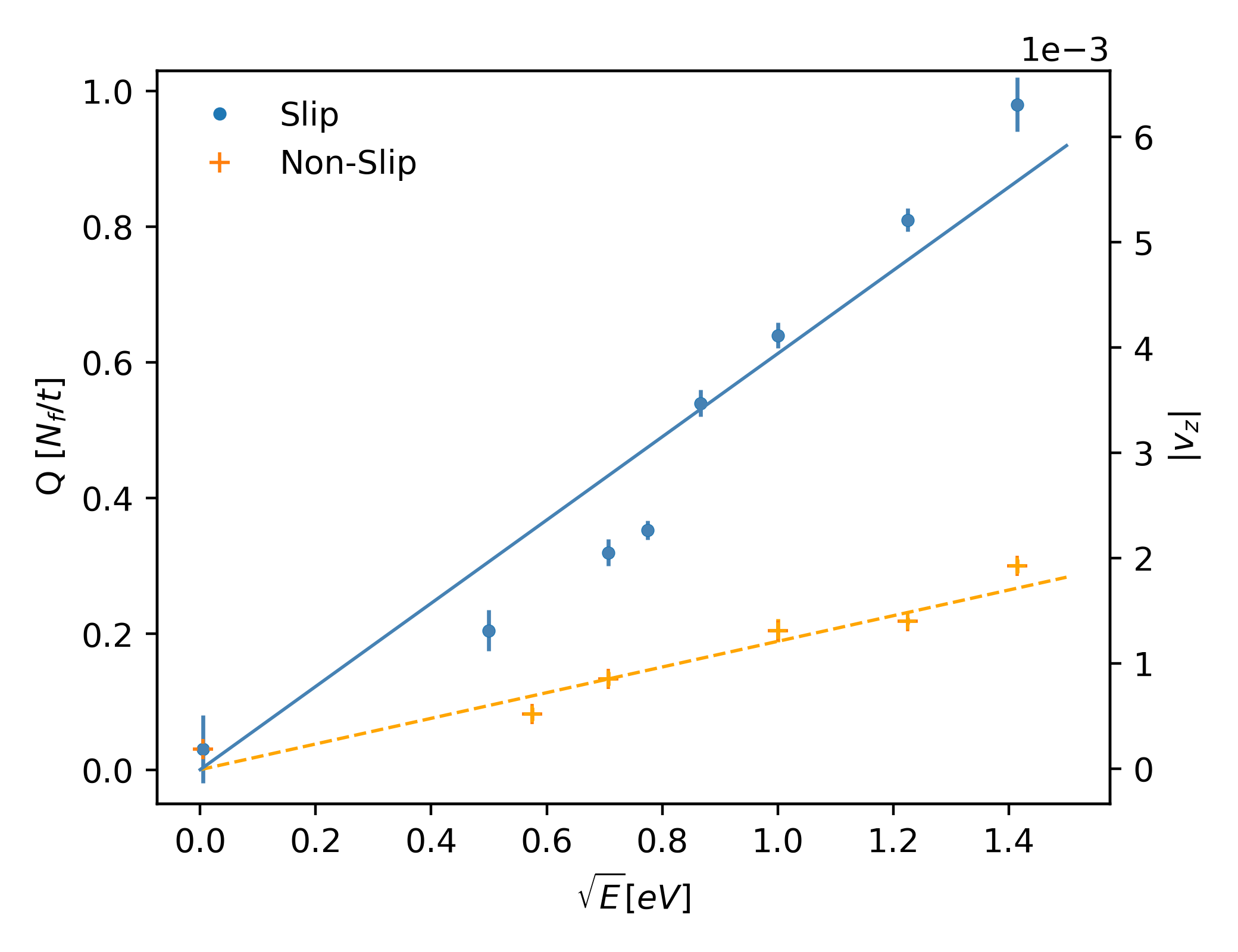}\\
    \small (a) 
     \\[\abovecaptionskip]
  \end{tabular}
  \begin{tabular}{@{}c@{}}
  \includegraphics[scale=0.7]{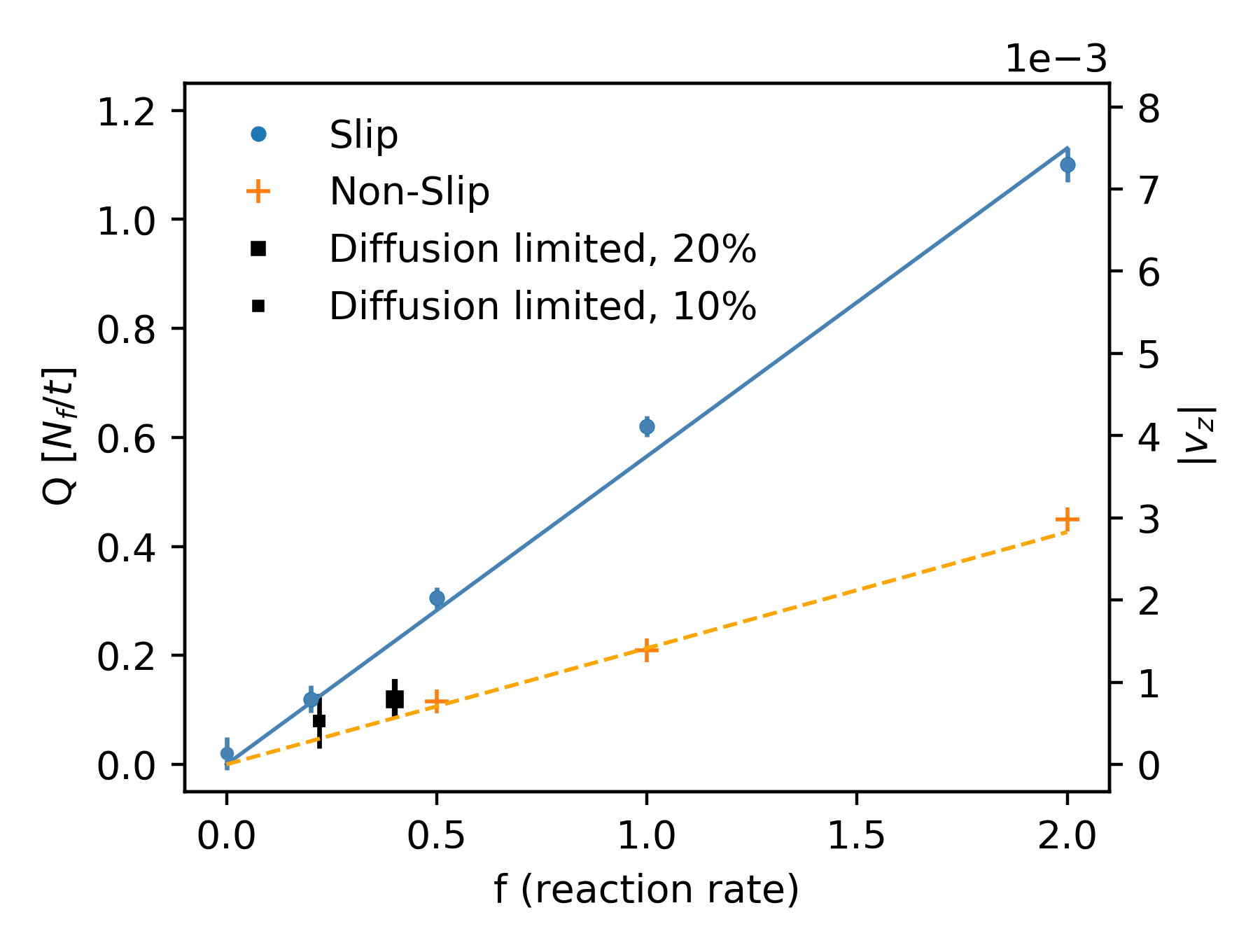}\\
  \small (b)
  \\[\abovecaptionskip]
  \end{tabular} 
  \caption{\label{fig:RvsEf}
  (a) Flow rate (Q, number of DPD fluid particles, $N_f$, per unit of time) as function of reaction energy with reaction rate $f=1$, for the case of a slip surface (pluses, fitted by dashed lines), and non-slip surface (circles, fitted by drawn line). The average flow velocity ($|v_z|$) is also shown in a secondary y-axis.
  (b) Flow rate (Q) and average flow velocity ($|v_z|$) as function of reaction frequency at \SI{1}{\electronvolt} for the slip (dashed line)  and non-slip (drawn line) surfaces.
 The two square symbols refer to the case of a 'diffusion limited' reaction at two molecular concentrations ($10\%$ and $20\%$).}
\end{figure}

The particle flow rate (Q) as function of the reaction energy and reaction rate ($f$) is shown in Fig. \ref{fig:RvsEf}a-b.
The flow rate is plotted for the two cases of a slip and non-slip boundary conditions (in what follows, we consider in both cases an inelastic surface and a thermostated fluid). 
Fig.~\ref{fig:RvsEf}a shows that the flow velocity varies as the square root of released energy, indicating that momentum is mainly a result of direct momentum transfer from reaction energy: in the scenario of thermo-osmosis the flow velocity would vary linearly with $E$.  In our simulations, the reaction rates were low enough to ensure that the flow velocity was linear in the reaction rate. This linear relation (Fig. \ref{fig:RvsEf}b) allows us to extrapolate our simulation results to realistic catalytic and electrochemical rates (which are few orders slower for water, as discussed below). 
\begin{table*}[h!]\centering \caption{Estimates of the transport in a cylindrical pump driven by surface reactions. The values in the table have been converted from DPD units to SI units for the case of water at {\bf 298} K.  The values refer to the geometry shown in Fig.~\ref{geometry} with non-slip and a thermostated wall.}
\begin{threeparttable}
\begin{tabular}{@{}llll@{}}
\toprule
Physical estimates & Rate $[ns^{-1}]$,$[s^{-1}nm^{-2}]$ & $L_z, R$ $[nm]$  & Flow Q[$\ce{H_2O}$/s], $v_z[mm/s]$ \\ \midrule
$f=0.1$ and \SI{1}{\electronvolt} & $33$, $3.6\cdot 10^9$ & 10.3, 2.6  & $1.05\cdot 10^9$, 4.5 \\
Double length  & $33$, $3.6\cdot 10^9$ & 20.6, 2.6 & $6.3 \cdot 10^8$, 2.7 \\
Double radius \& length & $72$, $3.6\cdot 10^9$ & 20.6, 5.2 & $1.87\cdot 10^9$, 1.97 \\
10\% fuel\tnote{a} & 4.0, $4.0\cdot 10^8$ & 10.3, 2.6 & $1.27\cdot 10^8$, 0.54 \\
10\% fuel, Reaction-limited rate\tnote{b} &  $0.017$, $1.8\cdot 10^6$ & 10.3, 2.6 &  $3.2\cdot 10^5$, $1.4\cdot 10^{-3}$ \\ \bottomrule
\end{tabular}
\begin{tablenotes}\footnotesize
\item[a] Diffusion-limited rate of electro-active fuel at 10\%. The reaction rate and the flux have been converted to values relevant for an aqueous solution, by scaling the diffusion coefficient and the viscosity of the DPD fluid to water.
\item[b] A representative reaction-limited rate equal to 0.25\%  of the diffusion-limited rate.
\end{tablenotes}
\end{threeparttable}
\label{table2}
\end{table*}

We also consider the case of a diffusion-limited rate of electro-active species in a short pump of a total length equals $L_z$. The short pump acts as a `gate', releasing fluid from one side ($-L_z/2$) to another side of the nano-channel ($L_z/2$). The pump switches-on flow by applying a sufficient over-potential on the active surface, so the maximum reaction rate is limited by active-species entering and diffusing through the channel towards the electro-active surface (electrode). Close to the active surface and in a reaction zone, active particles would undergo exothermic reactions with energy of \SI{1}{\electronvolt}. The reaction zone is set to 1.5 $r_c$, compatible with the typical range for electron-transfer in an aqueous electrochemical system (order of 1nm). 
The density of the active particles at the edge (-$L_z/2$) is kept constant to satisfy a Dirichlet boundary condition.     
Average rates of the stochastic reaction at two concentrations and their observed flow rates are plotted in Fig.~\ref{fig:RvsEf}b. 
We note that in addition to reactant diffusion, there is also advection. However, we found this non-zero Peclet effect to be relatively small $< 10\%$. 
For water, with a $10\%$ reactant concentration, a pump with a length of 10.3nm and a radius of 2.6nm produces a flow rate of $\sim 1.3$ water molecules per nano-second. The high flow rate obtained for the diffusion limited case shows that direct momentum transfer rate would be significant for smaller concentrations and slower reactions rate. Table \ref{table2} summarises the computed flow rates for an aqueous solution, as a function of the key geometric control parameters (tube length, tube diameter), and the reactant concentration. All results in Table~\ref{table2} were obtained for the non-slip, inelastic boundaries. 
The fluid flow velocity in water is predicted to be in the order of $1-10\mu m/s$ for catalytic rates which are usually kinetic limited. The velocity could be increased by a few order of magnitudes by tuning the size of the pump, the surface of the active area, and by adjusting the electro-chemical reaction rate. We can also compare the force from a reaction at a unit rate ($f=1$) to an effective force of a laminar flow in a cylinder, and roughly estimate the force in the pump of DPD fluid as function of size and rates. 
For  $L_z=16$, $R=4$ (in DPD units): $F_{avg}\sim 8L_z\nu Q/R^2 \sim 0.5$ where the kinematic viscosity of the fluid is $\nu \sim 0.3$. The effective force provides a crude but useful measure of how rates are influenced by the geometry. 
For instance, doubling the length $L_z$ to 20.6nm shows that rate is indeed reduced by around  50\% (Table \ref{table2}). 
As expected, the volumetric rate scales approximately as  $1/R^2$. 
We did not attempt to optimize the flow rates by systematically varying the tube geometry or the surface topography.

Other models of pumps based on temperature gradient show comparable velocities at a micro sized pump \cite{thermoRipoll}. 
High pumping rates ($\sim 1m/s$) theoretically can be obtained with high temperature gradients ($\sim 20-100^{\circ}K/nm$) at nano-sized pumps \cite{LiuThermo,thermosim}, but such extreme gradients may be difficult to achieve in experiments.
Experiments on diffusio-osmotic flow have found flow rates of $\sim 100$fl/min for a nano-channel with height of 163nm and width of $5\mu m$. 
If we roughly compare the diffusio-osmosis data to the case  presented here of a channel that is 3 orders of magnitude narrower, it would correspond to $\mathrm{Q}=5\cdot10^7\ce{H_2O}/s$ which is of the same order as the rates predicted on the basis of direct reactive momentum transfer. 

In summary, we have shown that heat release due to an exothermic surface reaction can drive fluid flow in a nano-channel.  
Energy from catalytic or electro-chemical reactions can generate momentum in the fluid which results in significant flow rates, provided that the active surface is placed in such a way that it favors uni-directional momentum transfer into the fluid. 
Fabrication of electrodes with such a directional preference could for instance be achieved using pulse electroplating or, possibly, nano-array printing techniques~\cite{arrayel}. 
The ease of design of the pump makes the momentum transfer mechanism attractive for applications and experimental studies.  
Further study of the momentum transfer mechanism is also required. In order to gain a better understanding of the factors that can affect the direct momentum transfer mechanism, more detailed, atomistic simulations would be needed. With such simulations, one could quantify the effect of thermalization at the wall and the influence of the topography of the active surface during heat release.

\section*{Acknowledgments}
 The work was funded by the EU  Horizon 2020 program through 766972-FET-OPEN-NANOPHLOW. 
 
\bibliography{nanopumpbib}
\bibliographystyle{unsrt} 
\end{document}


\maketitle

\section{Extrapolation to water viscosity}
\begin{figure}[h!]
\centering
\includegraphics[width=0.9\columnwidth]{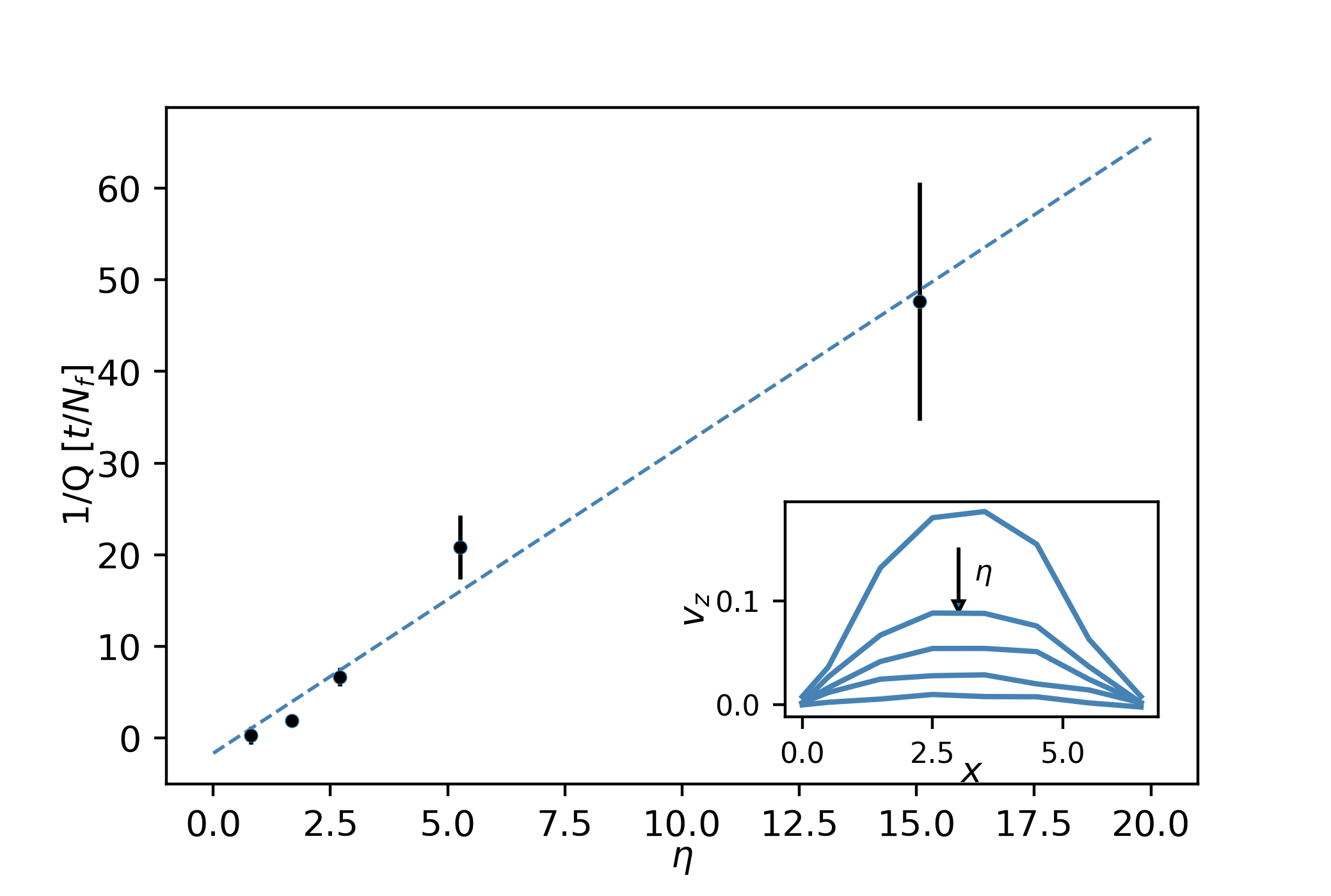}
\caption{Inverse flow rate as function of computed viscosity through Poiseuille flow and its linear fit (dashed line). The inset shows the $z$-velocity profiles along the $x$-direction. Simulations carried at higher viscosities were carried out by changing $\gamma$ and with $r_c$ equals 1.2 instead of 1.0 to achieve high viscosity with moderate values of $\gamma$.}
\label{flowvseta}
\end{figure}

In order to estimate the flow of an aqueous solvent, we must map the DPD results to fluid with the viscosity of water. 
This requires re-scaling the flow velocity, as  the viscosity of the DPD model proposed by Groot and Warren~\cite{groot1997dissipative} is $\sim 20$ times smaller than that of water, when expressed in the same units.
In the Stokes regime, the  the flow velocity is inversely proportional to the  viscosity. 
Such a dependence is indeed observed in Fig.~\ref{flowvseta}.

We estimated the viscosity of the DPD fluid by carrying out simulations of Poiseuille flow in a cylinder resulting from the application of a constant force per particle of $0.03N$ in the $z$ direction. 
We vary $\eta$ by changing $\gamma$ and $r_c$.
 The viscosity of the DPD fluid is estimated, assuming Poiseuille flow, in which case using  $\eta=\rho f_z R^2/(4v_{max})$.

\section{Pump Wall}
The pump-wall consists of `frozen' DPD particles at a (high) reduced density of $\sim 45$, thereby mimicking  a smooth and non-penetrable surface. 
To reproduce non-slip boundaries, we added a sinusoidal, longitudinal  corrugation to  the cylinder surface with an amplitude of 1/4 in reduced units and a period of $\pi/2$.  
We verified that this corrugation made the flow velocity vanish (less than 5\% from the maximum velocity) close to the wall in a body-force-driven flow through a cylinder. 
Alternatively, we can suppress friction with the wall (slip boundary conditions) by smoothing the boundaries. 

\begin{figure}[h!]
\centering
\includegraphics[width=0.85\columnwidth]{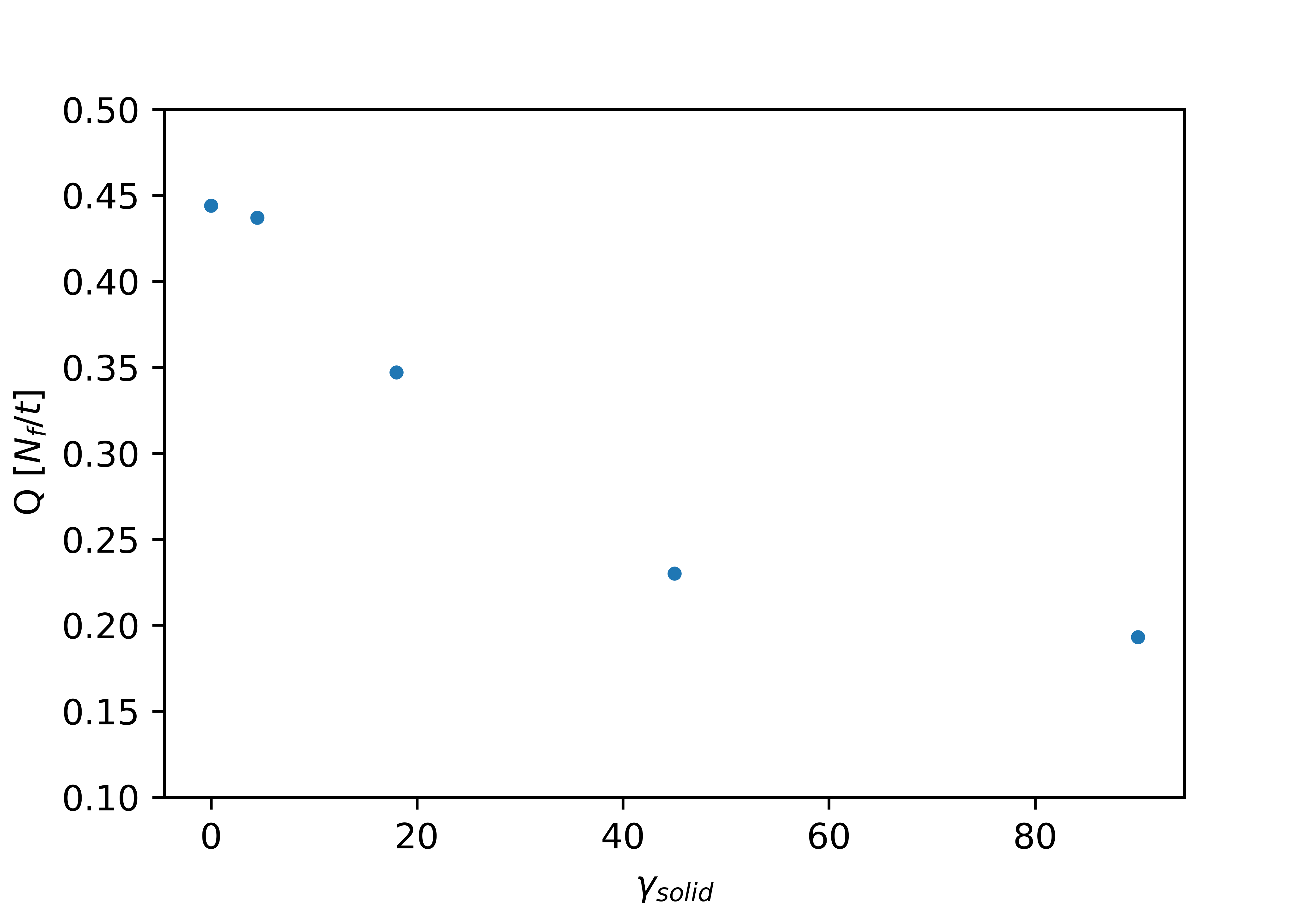}
\caption{Flow as function of $\gamma_{solid}$ for non-slip wall and a reaction rate $f$=1. In the simulations, we used a higher rate $f$=10, and rescaled the results back to $f$=1.
Energy of reaction is assumed to be 1eV.}
\label{flowg}
\end{figure}

The wall could be made to act as a thermostat,  by applying dissipative and random forces between wall and fluid particles with $\gamma_{solid}=10\cdot\gamma$. 
Depending on the geometry of the constriction in the channel, the thermalised case resulted in 25\%-50\% lower flow rate compared to the pure repulsive wall ($\gamma_{solid}=0$). 
For instance, Fig.~\ref{flowg} shows  the flow as a function of the thermalisation coefficient ($\gamma_{solid}$) in the case of a non-slip wall. 
We find that  the wall absorbs most energy on the  the active region, where reactions happens. 
Thermalising the cylindrical wall has a minor effect on the flow.
We stress that our model for reaction and flow at a solid-liquid interface is highly simplified: however, to obtain a more realistic description would require  fully atomistic simulations of the reaction dynamics and flow at a specific interface. As we are interested in generic effects, such simulations were beyond the scope of the present study.

All DPD calculations were carried out using the HOOMD-blue~\cite{hoomdpd} DPD code. 
The remainder of the code was written in PyCUDA.